\documentclass{emulateapj}
\usepackage{graphicx,natbib,amsmath,soul,xspace} 
\usepackage{CJKutf8}

\newcommand{\Msun}{$M_{\odot}$}

\newcommand{\Mstar}{$M_{*}$}

\newcommand{\ift}{$i^{\prime}$}

\newcommand{\gadget}{\textsc{gadget-3}\xspace}
\newcommand{\sunrise}{\textsc{sunrise}\xspace}

\begin{document}

\title{Merger signatures in the dynamics of star-forming gas}

\author{Chao-Ling Hung \begin{CJK*}{UTF8}{bsmi}(洪肇伶)\end{CJK*}\altaffilmark{1,2,3}}
\author{Christopher C. Hayward\altaffilmark{4,2}}
\author{Howard A. Smith\altaffilmark{2}}
\author{Matthew L. N. Ashby\altaffilmark{2}}
\author{Lauranne Lanz\altaffilmark{5}}
\author{Juan R. Mart{\'{\i}}nez-Galarza\altaffilmark{2}}
\author{D. B. Sanders\altaffilmark{1}}
\author{Andreas Zezas\altaffilmark{6,2}}

\affil{\altaffilmark{1} Institute for Astronomy, University of Hawaii, 2680 Woodlawn Dr., Honolulu, HI 96822, USA; chaoling.hung@gmail.com}
\affil{\altaffilmark{2} Harvard-Smithsonian Center for Astrophysics, 60 Garden St., Cambridge, MA 02138, USA}
\affil{\altaffilmark{3} Department of Astronomy, the University of Texas at Austin, 2515 Speedway Blvd., Austin, TX 78712, USA; Harlan J. Smith Fellow}
\affil{\altaffilmark{4} TAPIR, Mailcode 350-17, California Institute of Technology, 1200 E. California Blvd., Pasadena, CA 91125, USA}
\affil{\altaffilmark{5} Infrared Processing and Analysis Center, Caltech 100-22, Pasadena, CA 91125, USA}
\affil{\altaffilmark{6} University of Crete, Physics Department \& Institute of Theoretical \& Computational Physics, GR-710 03 Heraklion, Crete, Greece}


\begin{abstract}
The recent advent of integral field spectrographs and millimeter interferometers has revealed the internal dynamics of many hundreds of star-forming galaxies.
Spatially resolved kinematics have been used to determine the dynamical status of star-forming galaxies with ambiguous morphologies, and constrain the importance of galaxy interactions during the assembly of galaxies.
However, measuring the importance of interactions or galaxy merger rates requires knowledge of the systematics in kinematic diagnostics and the visible time with merger indicators.
We analyze the dynamics of star-forming gas in a set of binary merger hydrodynamic simulations with stellar mass ratios of 1:1 and 1:4.
We find that the evolution of kinematic asymmetries traced by star-forming gas mirrors morphological asymmetries derived from mock optical images, in which both merger indicators show the largest deviation from isolated disks during strong interaction phases.
Based on a series of simulations with various initial disk orientations, orbital parameters, gas fractions, and mass ratios, we find that the merger signatures are visible for $\sim0.2-0.4$ Gyr with kinematic merger indicators but can be approximately twice as long for equal-mass mergers of massive gas-rich disk galaxies designed to be analogs of $z\sim2-3$ submillimeter galaxies.
Merger signatures are most apparent after the second passage and before the black holes coalescence, but in some cases they persist up to several hundred Myr after coalescence.
About $20-60\%$ of the simulated galaxies are not identified as mergers during the strong interaction phase, implying that galaxies undergoing violent merging process do not necessarily exhibit highly asymmetric kinematics in their star-forming gas.
The lack of identifiable merger signatures in this population can lead to an underestimation of merger abundances in star-forming galaxies, and including them in samples of star-forming disks may bias the measurements of disk properties such as intrinsic velocity dispersion.
\end{abstract}


\section{Introduction}

The identification of galaxy mergers/interacting systems is critical to understand the role of interactions in the growth and assembly of galaxies.
Specifically, what is the relative importance between smooth or continuous accretion and discrete merger events in galaxy evolution \citep[e.g.,][]{Hopkins2006,Genel2008,Dekel2009}, and what roles do mergers play in triggering star formation and nuclear activity across cosmic time \citep[e.g.,][]{Engel2010,Hayward2013,Hung2013,Casey2014}?  
The vast majority of such constraints have been derived from large optical imaging surveys via measurements of galaxy pair fractions and identifying merger-induced disturbed structures \citep[e.g.,][]{Lin2004,Conselice2008,Lotz2011,Man2012}.
The observed abundance of mergers can then be used to test the predictions of galaxy evolution models after proper conversions from merger fractions to galaxy merger rates \citep{Kitzbichler2008,Lotz2008,Hopkins2010a}.

Measurements of merger fractions or the merger/disk nature of individual galaxies based on optical morphologies can be ambiguous.
Disturbed morphological structures like tidal tails and bridges are indisputable evidence of galaxy interactions \citep{Toomre1972,Barnes1992,Kim2002,Rothberg2004}, but these features often fade away at large distances due to surface brightness dimming \citep[e.g.,][]{Hibbard1997,Overzier2010,Hung2014}.
Some galaxy mergers exhibit highly clumpy, irregular star-forming regions that are visible in the rest-frame UV and optical wavelengths \citep[e.g.,][]{Miralles-Caballero2011,Petty2014}.
However, these features are also commonly seen in clumpy star-forming galaxies at intermediate ($z\sim0.1$) and high ($z\gtrsim1$) redshifts \citep[e.g.,][]{Elmegreen2004,Elmegreen2007,Fisher2014,Guo2015}, in which their star-forming clumps are formed through gravitational instabilities in highly unstable, turbulent disks \citep{Bournaud2007,Dekel2009a,Ceverino2010}.


Spectral lines from stars, neutral gas, molecular gas, and ionized gas of nearby galaxies \citep[e.g.,][]{de-Zeeuw2002,Helfer2003,Dicaire2008,Walter2008} trace galaxy dynamics out to different radii \citep[e.g.,][]{Yun1994,Aalto1999}, and in some cases they may reveal the evolution and interaction history of galaxies \citep{Davis2011}.
Emission lines from molecular gas and ionized gas are the most common tracers for a large sample of resolved galaxy kinematics out to $z\sim3$ \citep[e.g.,][]{Tacconi2006,Forster-Schreiber2009,Daddi2010,Gnerucci2011}, for which the gas traces the star-forming fuel and massive star forming regions.
In fact, kinematic structures traced by molecular and ionized gas have been used to reveal the dynamical status of galaxies independent of their visible morphologies \citep[e.g.,][and a review by \citealp{Glazebrook2013}]{Swinbank2006,Tacconi2006}; that is, whether galaxies display rotational patterns as expected for disks \citep[e.g.,][]{Daigle2006,Dicaire2008} or complicated kinematics as expected for mergers \citep{Mihos1998,Colina2005}. 
Recent large integral field spectrograph (IFS) surveys such as CALIFA \citep{Husemann2013}, SAMI \citep{Cortese2014}, MaNGA \citep{Law2015}, and KMOS$^{\rm 3D}$ \citep{Wisnioski2015} have significantly increased the sample of star-forming galaxies with resolved kinematics.
These observations are able to constrain merger abundances with respect to a wide range of galaxies' luminosities, stellar masses (\Mstar), and star formation rates (SFR), and complement to the studies based on optical imaging surveys. 

However, several complications attend kinematic diagnostics.
It has been demonstrated in both simulations and observations that gaseous disks are able to survive during the interaction between gas-rich systems or reform through accreting gas after two nuclei merge \citep[e.g.,][]{Downes1998,Barnes2002,Springel2005b,Hopkins2009,Ueda2014}.
These reformed disks can have \Mstar, SFR, and gas mass comparable to some of the $z\sim1-3$ star-forming disks \citep[e.g.,][]{Robertson2008}. 
Therefore, disk-like kinematics do not guarantee that the evolution history was quiescent.
Secondly, even during the earlier strong interaction stages, a small but significant fraction of mergers lacks the complicated kinematics expected from their disturbed morphology \citep[e.g.,][]{Mihos1998,Bellocchi2013}.
The contamination rates of mis-identified mergers/disks can be up to 50\% when classifying galaxies based solely on their resolved kinematics, and the results depend strongly on the interaction stage and the choice of kinematic classification schemes \citep{Hung2015}.

Comparisons between simulated and observed interacting galaxies have been used as a powerful tool to constrain detailed properties of mergers such as the initial encounter conditions \citep{Barnes2009,Privon2013}.
Although this detailed scrutiny for a large sample of galaxies is currently unattainable, mock observations based on hydrodynamic simulations can be used to study how the merger indicators evolve along the interaction sequence of different mass ratios, masses, and gas fractions \citep[e.g.,][]{Lotz2008,Lotz2010a,Lotz2010,Snyder2015}.
These studies also enable empirical calibrations of galaxy merger rates based on various morphological merger indicators.
Extensive work has been done exploring the kinematics of interacting galaxies and merger remnants using stellar populations as dynamical tracers \citep[e.g.,][]{Bendo2000,Jesseit2007,Naab2014,Stickley2014}, and some studies focus on the dynamics probed by the star-forming gas \citep[e.g.,][]{Robertson2008,Narayanan2009,Ceverino2012,Kassin2014}.
However, to date, there is a paucity of studies that systematically constrain the time intervals during which kinematic merger indicators are visible.

In this paper, we examine the evolution of kinematic merger indicators using a set of hydrodynamic simulations of binary mergers described in Section 2.
Specifically, we include merger simulations based on progenitor disks that are representative for local SDSS galaxies and $z\sim2-3$ submillimeter galaxies \citep{Lanz2014,Hayward2013}.
These simulations use widely employed SPH code \textsc{gadget}\xspace \citep{Springel2005a}, and their implementation of star formation and feedback are similar to many previous work \citep[e.g.,][]{Cox2006a,Robertson2006a}.
In Section 3, we detail the realization of mock kinematic maps and optical images.
The merger indicators used in this paper are described in Section 4.
We report our results in Section 5 and discuss their implications in Section 6.
We list our conclusions in Section 7.

\section{Simulated Galaxy Mergers}

We use a set of hydrodynamic simulations of galaxy mergers and isolated galaxies performed by \citet[][hereafter L14, also see \citealp{Hayward2014a,Martinez-Galarza2014}]{Lanz2014}.
These simulations are carried out using \gadget\ \citep{Springel2005a}, which computes gravitational interactions via a hierarchical tree method \citep{Barnes1986} and gas dynamics via smoothed-particle hydrodynamics\footnote{Although the traditional formulation of SPH can be inaccurate in some fluid mixing processes \citep[e.g.,][]{Agertz2007}, the type of idealized merger simulations performed here are insensitive to these limitations \citep{Hayward2014}.} \citep[SPH;][]{Gingold1977,Lucy1977}.
Each model galaxy contains a disk with stars and gas, a stellar bulge, a dark matter halo, and a supermassive black hole.
The gravitational softening lengths of the baryonic and dark matter particles are 100 pc and 400 pc, respectively.
Star formation and supernova feedback are implemented via the effective equation of state (EOS) method of the sub-resolution interstellar medium (ISM) model \citep{Springel2003}, and only gas particles with density higher than a threshold of $n\sim0.1$ cm$^{-3}$ are assumed to follow the effective EOS of this model. 
The instantaneous SFR of each gas particle is determined using a volumetric generalization of the Kennicutt-Schmidt relation, SFR $\propto \rho_{\rm gas}^N$ \citep{Schmidt1959,Kennicutt1998}, with $N=1.5$ \citep{Springel2003}.
Stellar winds are not included in these simulations.
L14 simulations also includes black hole accretion and AGN feedback models from \citet{Springel2005c}.

The simulations in L14 include a suite of galaxy mergers from four progenitor disks (named as M0, M1, M2, M3 in L14) that are representative of galaxies from the Sloan Digital Sky Survey (SDSS).
These progenitor disks are similar to G0, G1, G2, G3 in \citet{Jonsson2006a} and \citet{Cox2008} except that G0-G3 have slightly higher gas mass and \Mstar\ than M0-M3, and no supermassive black hole is included in G0-G3.
The disk component in M0, M1, M2, M3 has a central metallicity of 0.34, 0.5, 0.7, 1.6 $Z_{\odot}$ and it follows a metallicity gradient between $-0.04$ and $-0.06$ dex/kpc. 
Each gas particle undergoes self-enrichment at a rate determined by its SFR.
The new star particles formed during simulations are characterized by a formation time and a metallicity from their parent gas particles.
In this paper, we focus our morphological and kinematic analyses on the two most massive mergers from L14 (M3M3e and M3M2e, where ``e'' refers to one of the non-special disk orientations defined in \citealp{Cox2006}).
Simulated mergers M3M3e and M3M2e have a total \Mstar\ of $5.4 \times 10^{10}$ \Msun\ and  $8.44 \times 10^{10}$ \Msun, respectively, which are typical for  IFS surveys at $z\sim1-3$ \citep[e.g.,][]{Forster-Schreiber2009,Wisnioski2015}. 
Details of initial masses, numbers of SPH particles, gas fractions, disk orientations, and orbital parameters of M3M2e and M3M3e are summarized in Table~\ref{tab:param}.

In addition to the M3M3e and M3M2e simulations from L14, we perform variations based on these two simulations to explore the possible impacts from numerical resolutions, gas fractions, orbital parameters, and the choices of initial disk orientations.
We perform two high resolution runs with the particle numbers 5 and 10 times higher than the runs in L14.
The gas-rich versions of M3M3e and M3M2e are carried out by doubling the initial gas fraction of progenitor disks.
Motivated by the cosmological simulations of dark matter halos in \citet{Khochfar2006}, we test three different orbital parameters\footnote{\citet{Khochfar2006} show that almost half of major mergers with mass ratio $\leq4$ have near parabolic orbit ($e\sim1$) and the rest are dominated by bound orbits ($e<1$). In these three additional runs, we choose two near parabolic orbits ($e=0.95$) with different $r_p$ and the other orbit with smaller $e=0.8$. However, we note that \citet{Khochfar2006} use a dark matter-only simulation, and the orbital parameters of the dark matter halos may not necessarily correspond to the orbital parameters of the galaxies in the halos.} with various eccentricity ($e$) and pericentric distance ($r_p$).
Finally, we carry out additional M3M3 and M3M2 simulations with four special initial disk orientations defined in \citet{Cox2006}.
Detailed parameters used in these variations are summarized in Table~\ref{tab:param}.

Finally, to address how well kinematic analyses based on binary merger simulations (L14) apply to $z\sim1-3$ star-forming galaxies, we include two additional simulations from \citet[][hereafter H13]{Hayward2013} as a test case.
The b6b6e and b6b5e simulations from H13 have stellar mass ratios of 1:1 and 1:4, and the progenitor disks are scaled to $z=3$ based on the method described in \citet{Robertson2006}.
These two simulations are more gas rich than the M3M2e and M3M3e simulations in L14 (Table~\ref{tab:param}), but have physical properties (\Mstar, SFR, submillimeter flux densities, etc.) typical for $z\sim2-3$ submillimeter galaxies \citep[SMGs,][]{Hayward2011,Hayward2012,Michaowski2012}.
The gravitational softening length of dark matter is 200 pc in the H13 simulations.
Otherwise, the b6b6e and b6b5e simulations were configured identically to those in L14.

\begin{table*}
\centering
 \caption{Initial properties of L14, H13 simulations and their variations}
 \label{tab:param}
 \begin{tabular}{@{}clccccc}
 \hline
 \hline
Variations  &  Names   & Mass & Particle Number & Gas Fraction & Disk Orientations & Orbital Parameters \\  
            &          & [total, stellar, gas] & [halo, star, gas] & A, B\footnote{A and B denote the primary and secondary galaxies in the interacting system.} & A [$\theta$, $\phi$], B [$\theta$, $\phi$] & $e$,$r_p$ (kpc),$r_{init}$ (kpc)   \\
            &          & (10$^{10}$ \Msun) & ($10^5$) & & & \\
\hline 
L14         & M3M2e   & 167, 5.40, 1.13 & 2, 1, 0.6 &0.196, 0.242&30, 60, $-$30, 45& 0.95, 13.6, 250\\   
            & M3M3e   & 232, 8.44, 1.60 & 2.4, 1.2, 1 &0.196, 0.196&30, 60, $-$30, 45& 0.95, 13.6, 250\\
\hline
          &  M3M2e-res1 &  ---\footnote{When no values are listed in the columns, those parameters are identical to the M3M2e or M3M3e simulations.} & 10, 5, 3 & --- &--- &---     \\
Numerical &  M3M2e-res2 &  --- & 20, 10, 6 &--- &--- &---        \\
Resolution&  M3M3e-res1 &  --- & 12, 6, 5 &--- &--- &--- \\
          &  M3M3e-res2 &  --- & 24, 12, 10 &--- &--- &--- \\
\hline
             &  h  &  --- & --- & --- & 0, 0, 0, 0 & ---     \\
Disk         &  b  &  --- & --- & --- & 180, 0, 0, 0 & ---   \\
Orientations &  c  &  --- & --- & --- & 180, 0, 180, 0 & --- \\
             &  d  &  --- & --- & --- & 90, 0, 0, 0 & ---  \\
 \hline
Orbital    &  orb1  & ---& ---& ---& --- & 0.95, 6.8, 250   \\
Parameters &  orb2  & ---& ---& ---& --- & 0.95, 27.2, 250   \\
           &  orb3  & ---& ---& ---& --- & 0.8, 50,   \\
 \hline
Gas Fraction  & M3M2e (gas rich)  & ---&---&0.392, 0.484& ---&---\\ 
              & M3M3e (gas rich)  & ---&---&0.392, 0.392& ---&---\\
 \hline 
H13     & b6b5e & 575, 4.72, 18.88 & 1.2, 0.8, 0.8 & 0.8, 0.8 & 30, 60, $-$30, 45 & 0.95, 4.7, 49 \\
           & b6b6e & 912, 7.48, 29.92 & 1.2, 0.8, 0.8 & 0.8, 0.8 & 30, 60, $-$30, 45 & 0.95, 4.7, 49 \\

\hline
 
 \end{tabular}
\end{table*}

\section{Galaxy Morphology and Dynamics}

\subsection{Broadband Images}

We use the three-dimensional Monte Carlo radiative transfer code \sunrise \citep{Jonsson2006,Jonsson2010} to produce mock images of the simulated galaxies described in Section 2.
\sunrise determines the emission from stars and AGNs in the hydrodynamic simulations with SED templates \citep{Leitherer1999,Hopkins2007} and then performs radiative transfer calculations to account for the absorption, scattering, and re-emission by dust.
We adopt the same dust model as L14 (the Milky Way-type dust model of \citet{Draine2007}).
L14 discuss two possible treatments of the sub-resolution ISM structure during radiative transfer (i.e., whether dust mass is derived based on the diffuse gas content in the \citealp{Springel2003} model or the total gas content).
Here we adopt the conversion that dust mass is based on the diffuse gas content, which can better reproduce the  SEDs of the observed interacting galaxies (L14).

We derive optical morphological properties using the mock SDSS \ift-band ($\lambda_{\rm eff}=7439$\AA, $\Delta \lambda =1044$\AA) images produced from \sunrise.
Rest-frame optical wavelength is an ideal window to trace the disturbed structures induced by galaxy mergers because the emission is dominated by old stellar populations instead of the clumpy star-forming regions \citep[e.g.,][]{Abraham2003,Conselice2003,Lotz2004}, and it is available for a large sample of star-forming galaxies from $z\sim0$ out to $z\sim1-3$ \citep[e.g.,][]{van-der-Wel2012,Kartaltepe2014}.
The $\sim7000-8000$\AA\ regime is not severely affected by dust extinction except for extreme cases like ultraluminous and luminous infrared galaxies \citep[(U)LIRGs;][]{Haan2011,Hayward2012}.
No significant impacts from dust extinction are seen in our morphological analysis based on the \ift-band images throughout the M3M2e and M3M3e simulations.
We generate mock \ift-band images at 100 Myr intervals throughout the interaction sequence, and decrease the sampling steps to 20 Myr intervals during the strong interaction phase. 
For each snapshot, we obtain mock images from seven viewing angles sampled in a regular grid in spherical coordinate.

We then convert the mock images from \sunrise to images comparable to real observations.
First, we place our simulated galaxies at a distance of 100 Mpc, in which the plate scale of SDSS images ($0\farcs396$) corresponds to a physical size of $\sim200$ pc.
The observed number of counts is determined according to the surface brightness of galaxies at the assumed distance.
We then convolve the \sunrise images with the typical point spread function (PSF) of SDSS \ift\ observations ($\sim1\farcs3$), and add a noise frame extracted from the blank region in real SDSS \ift\ images. 
Examples of processed mock images from the M3M2e and M3M3e simulations are shown in the left panels of Figures~\ref{fig:stamp_g3g2e} and~\ref{fig:stamp_g3g3e}, respectively.

\begin{figure}
\centering
  \includegraphics[trim=12mm 0mm 0mm 15mm, clip=True, width=0.45\textwidth]{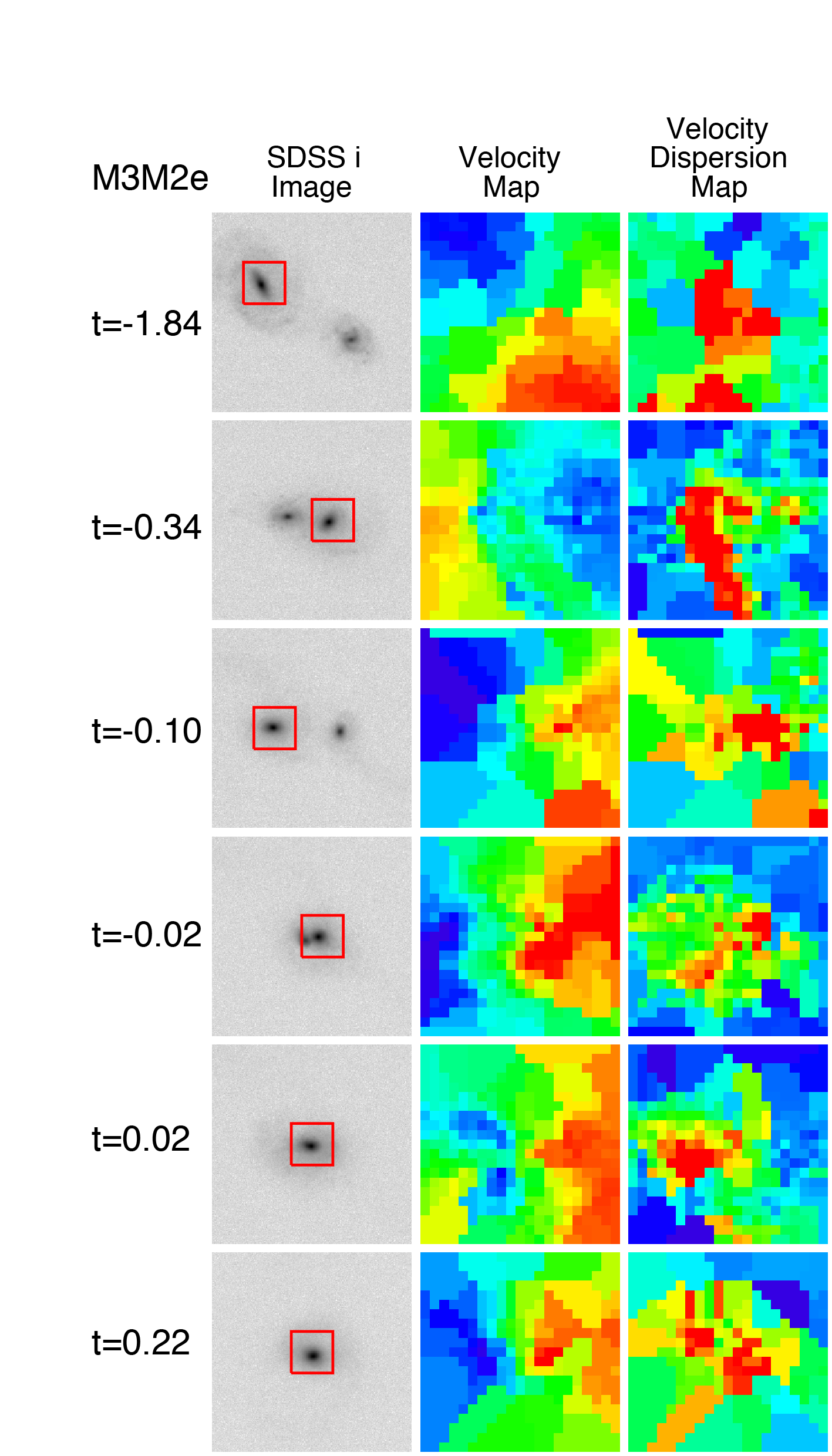} 
\caption{Example SDSS \ift-band images, velocity maps, and velocity dispersion maps of the M3M2e simulations from one viewing angle.
The time relative to the coalescence in each snapshot is shown at the left in units of Gyr (t=0 when the black holes coalescence).
The SDSS images in the left panels have the sizes of 50 kpc $\times$ 50 kpc.  
The velocity and velocity dispersion maps shown in the middle and right panels correspond to the kinematics in a subregion indicated by the 10 kpc $\times$ 10 kpc red boxes in the optical images.
The color scaling of kinematic maps is linear (red corresponds to larger values whereas blue corresponds to smaller values), but with arbitrary minimum and maximum values.
} 
\label{fig:stamp_g3g2e}
\end{figure}

\begin{figure}
\centering
  \includegraphics[trim=12mm 0mm 0mm 15mm, clip=True, width=0.45\textwidth]{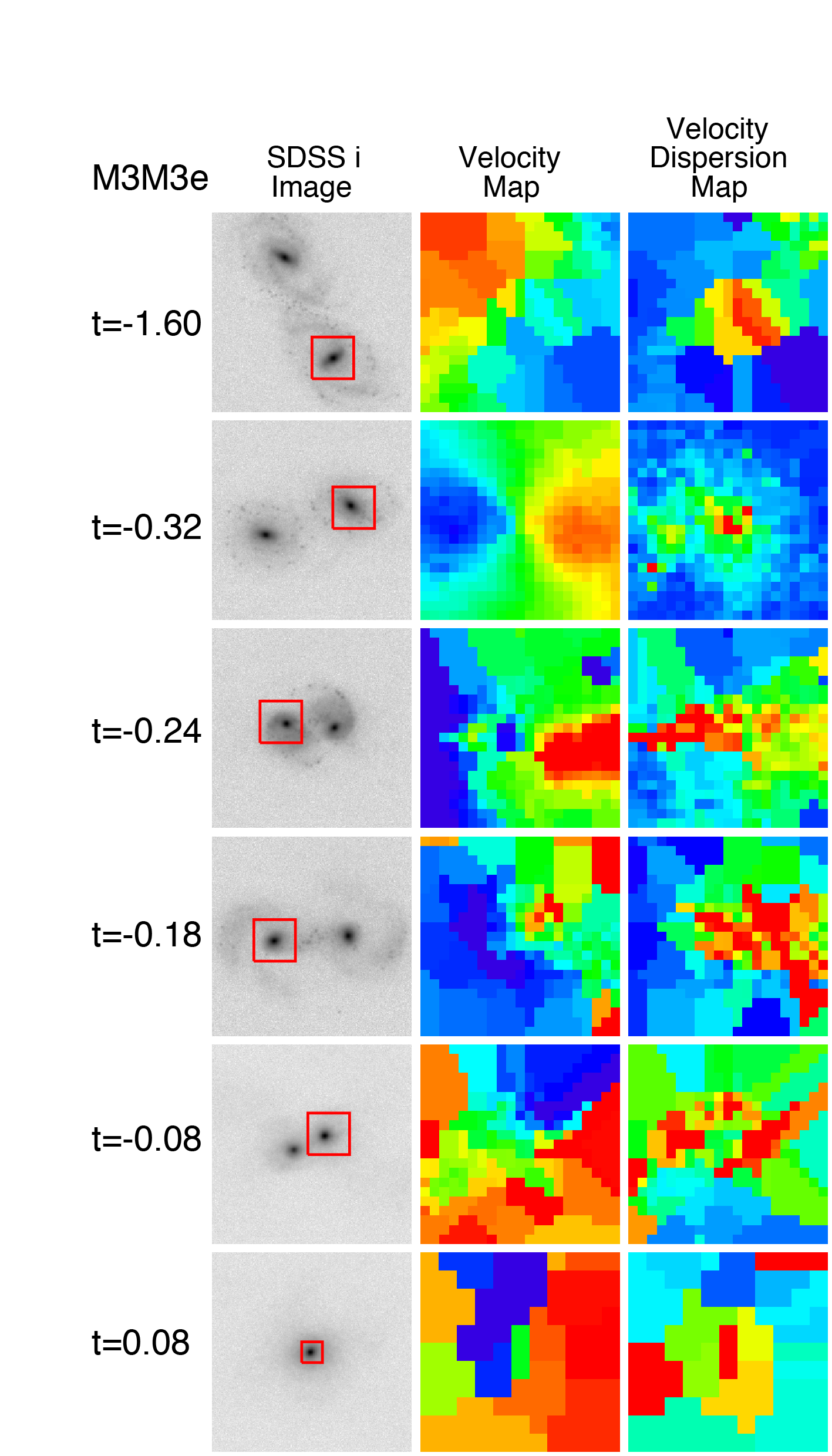} 
\caption{Same as Figure~\ref{fig:stamp_g3g2e} but for the M3M3e simulations.
The kinematic maps have the sizes of 10 kpc $\times$10 kpc (t=-0.08 Gyr and before) and 5 kpc $\times$5 kpc (t=0.08 Gyr).
} 
\label{fig:stamp_g3g3e}
\end{figure}

\subsection{Kinematic Maps}
As discussed in Section 1, emission lines from molecular gas and ionized gas are the most common tracers for a large sample of resolved galaxy kinematics at $z\sim0-3$.
Therefore, we focus our analysis on the kinematic properties derived from star-forming gas, and we discuss possible impacts using different dynamical tracers in Section 5.4.

We construct the kinematic maps based on the dynamical information from the SPH particles.
We select the subset of gas particles that have SFR $>0$ as a proxy of star-forming gas (where the gas density must be higher than a threshold of $n\sim0.1$ cm$^{-3}$) in the simulated galaxies.
In this simple approximation, possible impacts from dust are not included.
To convert particle-based information to kinematic maps, we make projected velocity and velocity dispersion maps from seven viewing angles that are consistent with \sunrise images.
In each viewing angle, we bin the gas particles into equally-spaced 500 pc $\times$ 500 pc bins (500 pc corresponds to $\sim$1\arcsec\ at the distance of 100 Mpc). 
The velocity and velocity dispersion in each pixel are then derived from the median and standard deviation of the gas particles weighted according to their SFR.
Finally, we adopt adaptive binning \citep{Cappellari2003} for the kinematic maps to ensure that each region (combined from $\geq$ 1 pixel) contains at least 10 star-forming gas particles.
Examples of velocity and velocity dispersion maps from M3M2e and M3M3e simulations are shown in the middle and right panels of Figures~\ref{fig:stamp_g3g2e} and~\ref{fig:stamp_g3g3e}, respectively.

\section{Merger Indicators}

\subsection{Kinematic Properties}
A common kinematic diagnostic of disks and mergers is the complexity of the galaxies' resolved kinematic properties, i.e., whether galaxies show ordered rotational patterns as expected for disk-like galaxies or chaotic patterns as expected for interacting systems.
Such identifications have been done via kinematic asymmetries \citep{Shapiro2008,Bellocchi2012}, visual inspections \citep[e.g.,][]{Flores2006,Epinat2012}, and visual comparisons with galaxy merger simulations \citep[e.g.,][]{Hammer2009}.

In this paper, we quantify how the degree of galaxies' kinematic maps deviate from a rotating disk using the kinematic asymmetries defined by \citet{Shapiro2008}, which is based on the higher-order moments kinematic coefficients of the velocity and velocity dispersion distributions derived using the kinemetry analysis \citep{Krajnovic2006}.
The line-of-sight velocity map or velocity dispersion map $K(a,\psi)$ can be divided into a series of elliptical rings (with semi-major axis $a$) as velocity or velocity dispersion profiles.
These profiles can then be described as an expansion of $N+1$ harmonic terms:
\begin{equation}
K(a,\psi)=A_0(a)+\sum\limits_{n=1}^N A_n(a) \sin{n\psi}+B_n(a) \cos{n\psi},
\end{equation}
where $\psi$ is the azimuthal angle.
\citet{Shapiro2008} quantify the level of deviation from an ideal disk by defining asymmetric measures of velocity and velocity dispersion fields as:
\begin{equation}
v_{asym} = \left< \frac{\sum\limits_{n=2}^5 k_{n,v}/4}{B_{1,v}} \right>_r,
\sigma_{asym} = \left< \frac{\sum\limits_{n=1}^5 k_{n,\sigma}/5}{B_{1,v}} \right>_r,\end{equation}
where $k_n=(A_n^2+B_n^2)^{1/2}$, the subscripts $v$ and $\sigma$ refer to the quantifies corresponding to the velocity and velocity dispersion maps, and $r$ refers to the average over all radii.
Finally, kinematic asymmetries, $K_{asym}$ is defined as $(v^2_{asym}+\sigma^2_{asym})^{1/2}$.

We measure $K_{asym}$ of all simulations from the velocity and velocity dispersion maps described in Section 3.2 using the {\tt IDL} routine {\tt Kinemetry}\footnote{http://davor.krajnovic.org/idl/} \citep{Krajnovic2006}.
We adopt the gas density peak position as the center of the kinematic maps, and then use {\tt Kinemetry} to find the best fit ellipse with position angle (PA) and the flattening factor ($Q=1-e$) at each radius step until more than 25\% of the data points along an ellipse are not present (the COVER parameter=0.75). 
The choice of this COVER parameter typically leads to an outer radius of $\sim$10 kpc during early interaction stages and $\sim$5 kpc during strong interaction and post-coalescence phases.
The evolution of $K_{asym}$ along the interaction sequence of M3M2e and M3M3e simulations is shown in the bottom panels of Figures~\ref{fig:sfr_asym_g3g2e} and~\ref{fig:sfr_asym_g3g3e}.
In general, only one galaxy in the interacting system (the one with higher central density) is included in the calculation when two galaxies are well-separated ($\gtrsim10$ kpc), and the evolution of $K_{asym}$ does not necessarily follow the same galaxy during the early interaction phases.
We also derive $K_{asym}$ in two additional cases following each galaxy in the interacting system, in which the centers of the kinematic maps are chosen at the positions of the supermassive black holes.
We note that {\tt Kinemetry} can fail to perform the elliptical fitting when the systems traced by the star-forming gas are too compact (e.g., $\lesssim$ 5 pixels across the galaxy), but typically less than 5\% of the data do not have $K_{asym}$ measurements in a given interaction sequence for this reason.

\subsection{Morphological Properties}

Various non-parametric statistics have been developed to quantify the irregularity of galaxy structure, and they can be used as indicators for possible disturbance due to galaxy mergers \citep{Conselice2000,Bershady2000,Conselice2003,Abraham2003,Lotz2004,Freeman2013}.
Extensive work has also been done to quantify the evolution of these parameters along the interaction sequence \citep[e.g.,][]{Conselice2006,Lotz2008,Lotz2010a,Lotz2010} and their robustness for nearby and distant galaxies \citep[e.g.,][]{Abraham1996,Overzier2010,Hung2014}.
In this paper, we quantify the morphological properties of the M3M2e and M3M3e simulations only to assist with the kinematic analysis, and refer the reader to the references listed above for detailed discussions of merger observability using morphological properties.

We measure the asymmetry parameter \citep[$A$;][]{Conselice2000} of galaxies in the M3M2e and M3M3e simulations from the mock SDSS \ift\ images.
We follow the definition of $A$ in \citet{Conselice2000}, in which it quantifies the deviation from $180^{\circ}$ rotational symmetry.
\begin{equation}
A=\sum_{i,j}\frac{\left|I(i,j)-I_{180}(i,j)\right|}{\left|I(i,j)\right|}-\sum_{i,j}\frac{\left|B(i,j)-B_{180}(i,j)\right|}{\left|I(i,j)\right|},
\end{equation}
where $I$ and $I_{180}$ is the galaxy image and its $180^{\circ}$ rotated version, and $B$ and $B_{180}$ represent the background and its $180^{\circ}$ rotation.
$A$ is often significantly enhanced relative to elliptical or spiral galaxies in the presence of multiple bright components and extremely irregular structure.
The merger simulations in \citet{Lotz2008,Lotz2010a,Lotz2010} have also demonstrated that $A$ is most sensitive to interacting galaxies during the strong interaction phases before the final coalescence and in some cases during the first passage as well.

To derive $A$, we first use \texttt{SExtractor} \citep{Bertin1996} to identify galaxies in each mock \ift-band images.
The de-blending parameters have been chosen so that the interacting systems are identified as one galaxy when the projected distance between two nuclei is smaller than $\sim5-10$ kpc.
When more than one object is detected in the images, we mask out the detections other than the brightest galaxy and we refill the masked regions with nearby sky.
In this case, most of the identified regions along the interaction sequence for deriving $A$ are consistent with the kinematic measurements.
We apply a ``quasi-Petrosian'' method \citep{Abraham2007} to define the Petrosian radius ($r_p$) as the effective radius at the isophotal threshold of 0.2 and we define the center of galaxies as where $A$ is minimized \citep{Conselice2000}.
Finally, $A$ is derived by summing over all pixels within 1.5 $r_p$ (Equation 1).
The evolution of $A$ along the interaction sequence of M3M2e and M3M3e simulations is shown in the middle panels of Figures~\ref{fig:sfr_asym_g3g2e} and~\ref{fig:sfr_asym_g3g3e}.

\begin{figure}
\centering
  \includegraphics[width=0.5\textwidth]{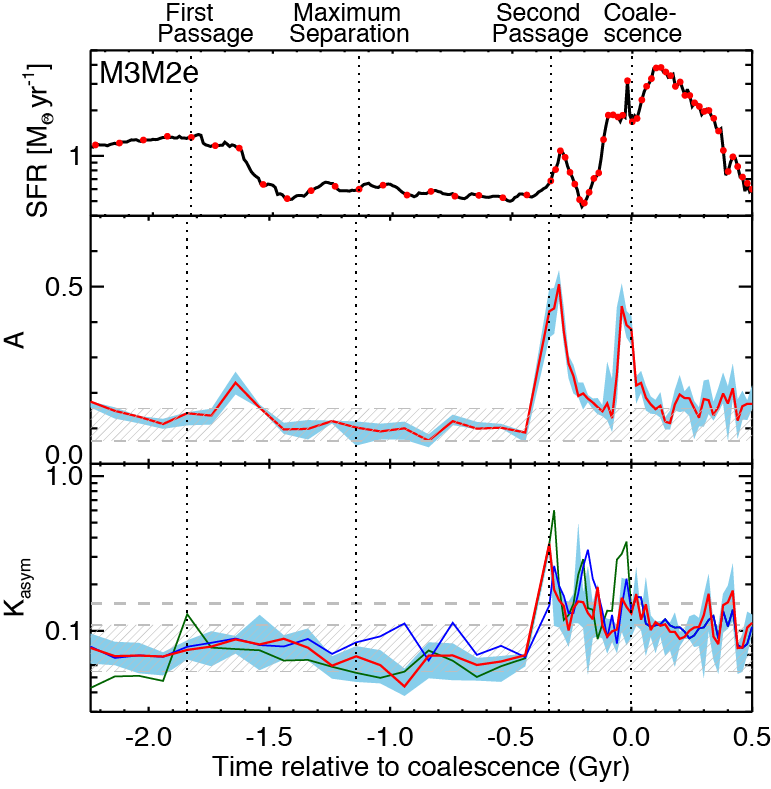} 
\caption{SFR, $A$, and $K_{asym}$ during the interaction sequence of the M3M2e simulations.
The dotted lines indicate the stages of interaction: first passage, maximum separation, second passage, and final coalescence.
The red dots overlaid on the SFR curve represent the snapshots that we derive $A$ and $K_{asym}$ from optical images and kinematic maps.
In the middle and bottom panels, the red solid lines show the median $A$ and $K_{asym}$ curve of the values from seven viewing angles.
The blue area indicates the distribution of 70\% of the data (values from five viewing angles near the median values).
The blue and green solid lines indicate the median $K_{asym}$ curves of the major and minor companion, in which the center of kinematic maps are defined as the positions of supermassive black holes.
The gray shaded area indicate the distribution of 70\% of the values from the isolated M3 galaxy simulations.
The thick gray dashed line shows a merger/disk classification criterion of $K_{asym}=0.15$, which is defined based on the value higher than 95\% of the isolated M3 galaxy simulations.
} 
\label{fig:sfr_asym_g3g2e}
\end{figure}

\begin{figure}
\centering
  \includegraphics[width=0.5\textwidth]{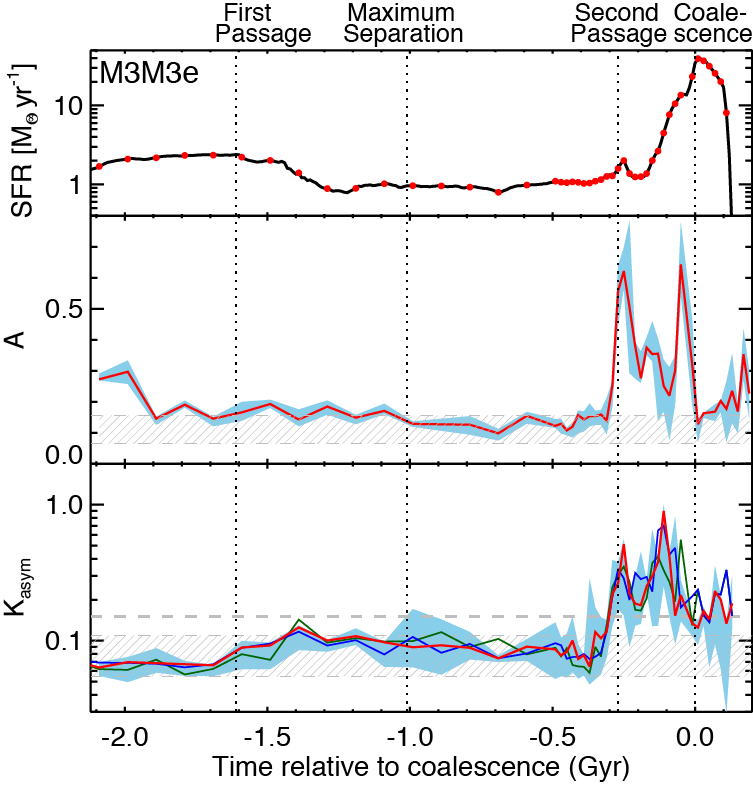} 
\caption{Same as Figure~\ref{fig:sfr_asym_g3g2e} but for the M3M3e simulations.
} 
\label{fig:sfr_asym_g3g3e}
\end{figure}

\begin{figure}
\centering
  \includegraphics[width=0.5\textwidth]{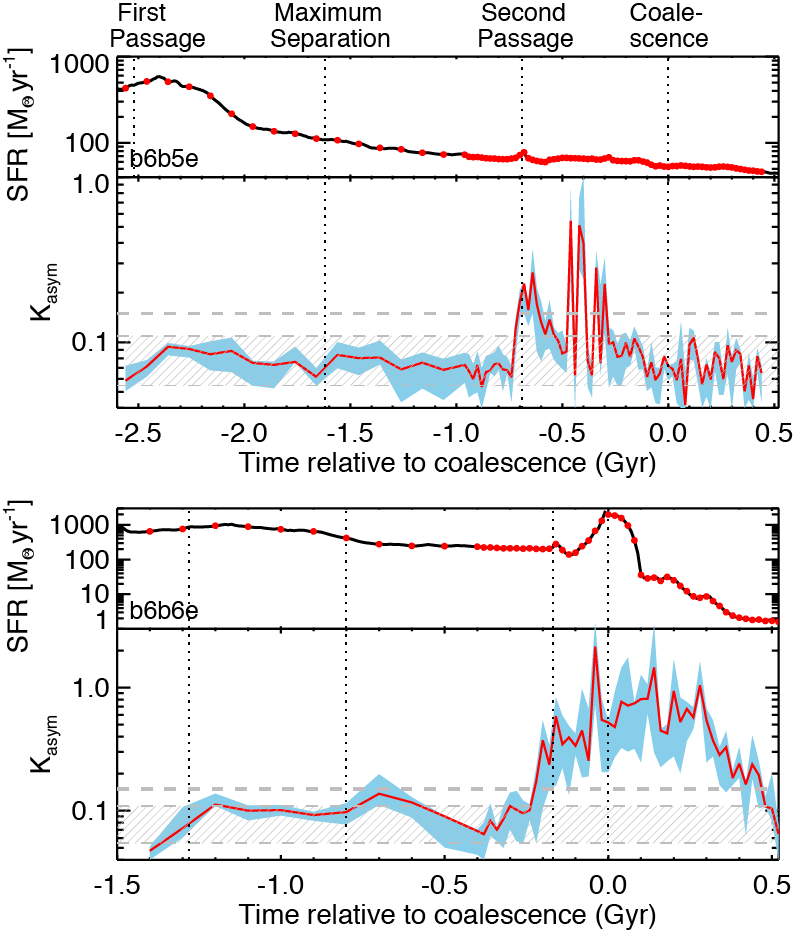} 
\caption{SFR and $K_{asym}$ during the interaction sequence of the b6b5e and b6b6e simulations.
The dotted lines indicate the stages of interaction: first passage, maximum separation, second passage, and final coalescence.
The red dots overlaid on the SFR curves represent the snapshots that we derive $K_{asym}$ from kinematic maps.
The red solid lines show the median $K_{asym}$ curves of the values from seven viewing angles.
The blue area indicates the distribution of 70\% of the data (values from five viewing angles near the median values).
The gray shaded area indicate the distribution of 70\% of the values from the isolated M3 galaxy simulations.
The thick gray dashed line shows a merger/disk classification criterion of $K_{asym}=0.15$.
} 
\label{fig:smg}
\end{figure}

\section{Results}
\subsection{Merger indicators along the interaction sequence}
Figures~\ref{fig:sfr_asym_g3g2e} and~\ref{fig:sfr_asym_g3g3e} show the evolution of SFR, $A$, and $K_{asym}$ along the interaction sequence of the M3M2e and M3M3e simulations.
The distribution of $A$ and $K_{asym}$ from isolated M3 simulations with various viewing angles and time are indicated in gray shaded area.
In both M3M2e and M3M3e simulations, $A$ is significantly enhanced only after the second passage of galaxies and before coalescence.
During this strong interaction phase, individual galaxies display large scale tidal features and lead to high $A$ even when two galaxies can still be resolved.
When the two nuclei are close enough ($\lesssim5-10$ kpc) to be considered as one system, the multiple bright components can also result in higher values of $A$.
The enhancement of $A$ during the strong interaction phases are consistent with the results of G3G3P and G3G2P simulations in \citet{Lotz2008,Lotz2010}, which use similar progenitor galaxies and orbital parameters but different initial disk orientations.
Although a small fraction of the data in \citet{Lotz2010} have elevated $A$ during the first passage, no significant enhancement is seen in our M3M2e and M3M3e simulations.

The evolution of $K_{asym}$ approximately tracks $A$ before the coalescence phase in both M3M2e and M3M3e simulations.
The low $K_{asym}$ during the early interacting stages demonstrates that within individual galaxies, only minimal disturbance is seen in the kinematic structures traced by star-forming gas.
Although galaxy interactions may begin to affect the SFR and metallicity of individual galaxies during the early phase of interaction \citep[e.g.][]{Scudder2012,Moreno2015}, this impact may not necessarily reflect on the irregularity in galaxy kinematics.
The lack of detectable enhancement in $K_{asym}$ is the case for each galaxy in the interacting systems.
We derive $K_{asym}$ in two additional cases following two individual galaxies in which the center of the kinematic maps coincide with the positions of supermassive black holes (blue and green solid lines in the bottom panels of Figures~\ref{fig:sfr_asym_g3g2e} and~\ref{fig:sfr_asym_g3g3e}).
The resulting median $K_{asym}$ curves show similar trends with the kinematic maps centered at the gas density peak.

From right after the second passage through the coalescence phases, $K_{asym}$ show significant deviations from the isolated M3 simulations. 
Most of the snapshots during this strong interaction phase display highly disturbed structure in both velocity and velocity dispersion maps, in which the kinematic structures are dominated by the bulk motion of two nuclei and the merger-induced gas flows.
The oscillations of $K_{asym}$ between second passage and coalescence reflect the projected distance between two nuclei; stronger disturbances are measured when two nuclei approach each other whereas such disturbances decrease as the two nuclei recede from each other.
After the two nuclei merge, a gaseous disk survives in the M3M2e simulations and its $K_{asym}$ decreases to the level of isolated M3 simulations.
However, no such structure is formed in the M3M3e simulations, and most of the gas has funneled to the galaxy center and been consumed by the starbursts within $\sim100-200$ Myr.
The $K_{asym}$ of the M3M3e simulations remain slightly enhanced after the coalescence phase for $\sim100$ Myr until the star-forming gas is exhausted (SFR $\lesssim0.5$ \Msun\,yr$^{-1}$) and kinematics is no longer traced.

In Figure~\ref{fig:smg}, we show the evolution of SFR and $K_{asym}$ in the b6b5e and b6b6e simulations, which are binary mergers of SMG-type progenitors as described in Section 2. 
Prior to the coalescence phases, the b6b5e and b6b6e simulations have significantly higher SFR than the M3M2e and M3M3e simulations because the SMG-type progenitor disks are more gas rich and have higher gas densities.
Despite these differences, the evolution of $K_{asym}$ in b6b5e and b6b6e shows a similar trend as M3M2e and M3M3e.
For instance, $K_{asym}$ only begins to elevate significantly after the second passage.
A key difference seen between the equal mass mergers M3M3e and b6b6e is that $K_{asym}$ of b6b6e is elevated for $\sim400$ Myr after black hole coalescence.
This prolonged disturbance in the dynamics of star-forming gas is visible due to a more gradual decline in SFR after coalescence (i.e. it only takes $\sim$ 0.25 Gyr for M3M3e to reach a SFR that is 0.01\% of its peak SFR after black hole coalescence, whereas it takes $\sim$1 Gyr for b6b6e to reach 0.01\% of its peak SFR).

\begin{table*}
\centering
 \caption{Merger observable time defined based on $K_{asym}$}
 \label{tab:timescale}
 \begin{tabular}{@{}clclc}
 \hline
 \hline
            &   \multicolumn{2}{c}{Mass Ratio $=$ 1:4} &   \multicolumn{2}{c}{Mass Ratio $=$ 1:1} \\
 Variations\footnote{Detailed parameters used in these variations are listed in Table~\ref{tab:param}.} & Configurations & Observable Time (Gyr) &  Configurations & Observable Time (Gyr) \\
 \hline
 L14              & M3M2e & 0.22$\pm$0.04  & M3M3e & 0.36$\pm$0.06 \\
 \hline
Numerical   & M3M2e-res1 & 0.28$\pm$0.04 & M3M3e-res1 & 0.26$\pm$0.03  \\
Resolution & M3M2e-res2 & 0.32$\pm$0.04   & M3M3e-res2 & 0.42$\pm$0.02 \\
 \hline
			            & M3M2h & 0.34$\pm$0.03  & M3M3h & 0.26$\pm$0.05 \\              
Disk                 & M3M2b  & 0.24$\pm$0.03 & M3M3b & 0.34$\pm$0.04 \\ 
Orientations     & M3M2c  & 0.20$\pm$0.05  & M3M3c & 0.30$\pm$0.05  \\ 
                        & M3M2d  & 0.26$\pm$0.03  & M3M3d & 0.34$\pm$0.03 \\ 
 \hline 
			            & M3M2e-orb1  & 0.20$\pm$0.06  & M3M3e-orb1 & 0.32$\pm$0.06 \\              
Orbital             & M3M2e-orb2  & 0.26$\pm$0.04  & M3M3e-orb2 & 0.48$\pm$0.03  \\ 
Parameters      & M3M2e-orb3  & 0.20$\pm$0.05  & M3M3e-orb3 & 0.30$\pm$0.06 \\ 
                          
 \hline             
Gas Fraction & M3M2e (gas rich) &  0.22$\pm$0.03    & M3M3e (gas rich) & 0.34$\pm$0.03 \\   
\hline 
H13 & b6b5e &  0.20$\pm$0.03   & b6b6e &0.72$\pm$0.07 \\   
\hline
 \end{tabular}
\end{table*}

\subsection{Merger observable time and probability with kinematic indicators}
We derive the merger observable time (i.e. the time duration that merger signatures are detectable, hereafter MOT) using the median $K_{asym}$ curves (e.g., the bottom panels of Figures~\ref{fig:sfr_asym_g3g2e} and~\ref{fig:sfr_asym_g3g3e}).
We define the criterion of a galaxy to be classified as a merger when its $K_{asym}$ is significantly enhanced, and here we use a threshold of $K_{asym}=0.15$ (a value higher than 95\% of galaxies from the isolated M3 simulations).
We note that this threshold is comparable to the one defined by \citet{Bellocchi2012} but considerably lower than the criteria used by \citet{Shapiro2008}.
Since our criteria are defined using simulations of the progenitor disk followed with the same kinematic mapping as the merger simulations, any enhancement in $K_{asym}$ can be attributed as a result of interactions.
The derived MOT with $K_{asym}>0.15$ are 0.22 and 0.36 Gyr for the M3M2e and M3M3e simulations (Table~\ref{tab:timescale}).
The uncertainties are derived based on the 1 $\sigma$ distributions of the median $K_{asym}$ curves (the blue shaded area in Figures~\ref{fig:sfr_asym_g3g2e} and~\ref{fig:sfr_asym_g3g3e}).
Results based on different numerical resolutions typically differ within $\pm$0.1 Gyr.

Since we attribute the main source of uncertainty as the variation due to the viewing angles, it is important to examine whether our choice of seven viewing angles are truly representative to the typical variation in $K_{asym}$.
We derive $K_{asym}$ of 70 viewing angles for two snapshots of M3M2e with one in the early interaction stage and the other close to the coalescence.
We find that in both snapshots, the 1 $\sigma$ distribution of the data points from 7 viewing angles span a range similar to the distribution derived based on 70 viewing angles.
Another concern is whether a time step of 100 Myr is sufficient to trace the variation during the early interaction phases.
We have increased the sampling in timestep to 20 Myr before the second passage, and the MOT only increases 20 Myr for the M3M2e simulations and does not change for the M3M3e simulations.

We explore the dependence of MOTs on initial conditions of galaxy merger simulations.
Specifically, we focus on whether the choices of gas fractions, orbital parameters, and initial disk orientations may have significant impacts (Table~\ref{tab:timescale}).
The gas rich runs of M3M2e and M3M3e with doubling the initial gas fraction can lead to molecular gas fraction comparable to local LIRGs or ULIRGs type objects \citep[e.g.,][]{Sanders1991}, yet their MOTs remain similar to the original runs.
The results from various orbital parameters span a wider range ($0.2-0.48$ Gyr), in which ``orb2 ($e=0.95$, $r_p=27.2)$'' have larger MOTs due to its $\sim$ twice longer duration between second passage and coalescence.
We also carry out simulations with four special initial disk orientations, and these variations lead to observable time of $0.20-0.36$ Gyr.
In all variations based on L14 simulations, merger signatures in $K_{asym}$ are most visible during the strong interaction phase and only visible for $\lesssim$ 100 Myr after black holes coalescence regardless the availability of star-forming gas.
The equal-mass merger simulation with SMG-type progenitors (b6b6e) has doubled MOT compared to M3M3e, in which merger signatures are visible for $\sim0.4$ Gyr during the post-coalescence phase until its SFR decreases to $\sim0.5$ \Msun\,yr$^{-1}$.

The merger/disk classification criteria and the time when the $K_{asym}$ curves end may introduce additional systematics to MOTs.
For example, if we apply a lower classification threshold, e.g., $K_{asym}=0.11$ (a value higher than 68\% of galaxies from the isolated M3 simulations), then the MOTs of the M3M2e and M3M3e simulations increase to 0.48 and 0.56 Gyr, respectively.
On the other hand, our kinematic analysis stops when SFRs of merger remnants are $\lesssim0.5$ \Msun\,yr$^{-1}$, where no sufficient gas particles are available to make kinematic maps with a even lower SFR.
In the case that the disk structure is completely destroyed during the interaction, $K_{asym}$ remains elevated after black hole coalescence and MOTs may be sensitive to the choice of SFR limits to derive $K_{asym}$.
However, MOTs of M3M3e and b6b6e do not change significantly when varying SFR limits to several \Msun\,yr$^{-1}$.

When treating the simulated galaxies at each snapshot and viewing angle as individual systems, we can quantify the observable merger fractions as a function of interaction stages.
Figure~\ref{fig:merger_kin} shows the fraction of simulated galaxies classified as mergers using the criterion $K_{asym}\geq0.15$ for simulations with five different initial disk orientations.
Before coalescence, the derived merger fractions of all the M3M2 simulations agree within $\sim20-40\%$ and the M3M3e simulations are typically higher compared to the M3M2e counterparts in all interaction stages. 
As expected based on the results shown in Section 5.1, the derived merger fractions past the coalescence phases show larger scatters as a result of different remnants in these simulations.
These merger fractions are comparable to the results in \citet{Hung2015} when they use the classification scheme in \citet{Shapiro2008} and systematically lower by $\sim50\%$ when they use the classification scheme in \citet{Bellocchi2012}.

\begin{figure}
\centering
  \includegraphics[width=0.5\textwidth]{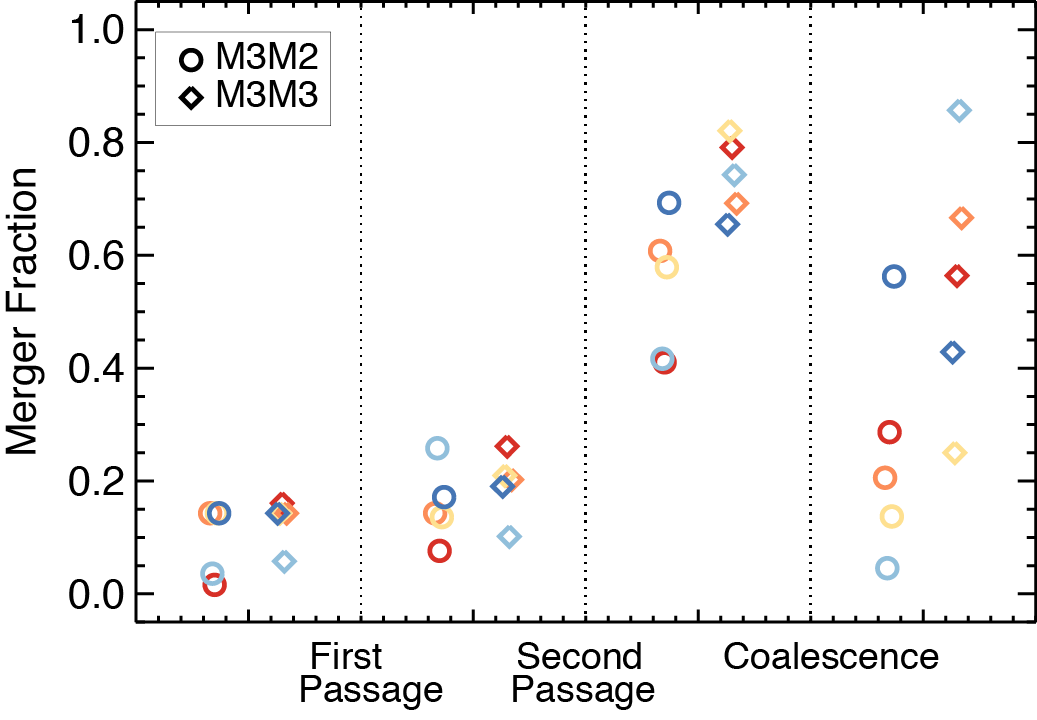} 
\caption{Merger fractions derived based on $K_{asym}$ for merger simulations with various initial disk orientations at four interaction stages: before first passage, after first and before second passage, after second passage and before coalescence, and after coalescence.
The circles are M3M2 simulations and the diamonds are M3M3 simulations.
Different colors indicate different initial disk orientations: red (e), orange (b), light orange (c), light blue (d), and blue (h).
} 
\label{fig:merger_kin}
\end{figure}

\subsection{Dependence on spatial resolution}
Spatial resolution is critical for accurately deriving galaxy kinematic properties.
For example, \citet{Goncalves2010} find that the merger fraction of Lyman Break Analogs at $z\sim0.2$ decreases by a factor of two (from $\sim70\%$ to $\sim38\%$) when artificially redshifting the sample to $\sim2.2$, where the spatial resolution is 10 times worse in the redshifted datacubes compared to the original ones.
The kinematic measurements in this paper are derived using kinematic maps with a spatial resolution of 0.5 kpc, which can be achieved in seeing-limited observations of local galaxies \citep[e.g.,][]{Husemann2013} and adaptive optics-assisted observations out to $z\sim0.4$ \citep[e.g.,][]{Goncalves2010}.
However, typical IFS surveys of $z\sim1-3$ galaxies often have spatial resolution of $\gtrsim$ 1 kpc \citep[e.g.,][]{Law2009} except for lensed galaxies \citep[e.g.,][]{Yuan2011,Livermore2015}.

We examine how our kinematic measurements of the M3M2e simulations vary if the spatial resolutions of kinematic maps decreases from 0.5 kpc to 1 kpc.
We create the kinematic maps following the description in Section 3.2 but replace the 500 pc $\times$ 500 pc grids with the 1 kpc $\times$ 1 kpc grids.
To ensure a consistent classification as discussed in Section 5.2, we also create low-resolution kinematic maps for the isolated M3 simulations and re-define the merger classification threshold for the low resolution maps as $K_{asym} \geq 0.192$ (higher than 95\% of the galaxies derived from the isolated M3 simulations).
The MOT derived from the median $K_{asym}$ curve decreases from 0.22$\pm$0.04 Gyr with 0.5 kpc resolutions to only 0.14$\pm$0.04 Gyr with 1 kpc resolutions.
This result demonstrates that with worse spatial resolution, the contrast between disturbed kinematics and comparison disks becomes smaller and thus the merger observable time becomes shorter.

\subsection{Gas kinematics versus stellar kinematics}

So far, our analyses have focused on galaxy kinematics traced by star-forming gas.
However, the flows of stars and gas during galaxy interactions may diverge in the presence of large-scale shocks \citep[e.g.,][]{Barnes1991,Barrera-Ballesteros2015}.
It is thus intriguing to quantify how the kinematic merger indicator, $K_{asym}$, may depend on which observational tracers are used along the interaction sequence.
We create the stellar kinematic maps following the procedures described in Section 3.2 using all of the stellar particles in the simulations.
The center of the kinematic maps are chosen as the positions of the supermassive black holes.
The velocity and velocity dispersion in each initial bins are determined as the median and standard deviation of all stellar particles weighted according to their masses.

Figure~\ref{fig:kasym_sdb} shows the median $K_{asym}$ curves of the M3M2e and M3M3e simulations derived based on all star particles until the end of our simulations ($\sim1.5$ Gyr after coalescence).
Although the star particles in general trace galaxy structure to larger radii than the star-forming gas throughout the interaction, the median $K_{asym}$ curve traced by stars progresses similarly as the curve traced by star-forming gas in both M3M2e and M3M3e simulations.
In both simulations, $K_{asym}$ does not increase significantly until second passage but the enhancement of $K_{asym}$ lasts through the entire strong interaction phase.
After coalescence, the remnant of the M3M2e simulations exhibits a rotational pattern, and its $K_{asym}$ reaches a lower, stable value than the $K_{asym}$ during strong interaction phase.
The remnant of  the M3M3e simulations still show highly disturbed kinematic structure, and its $K_{asym}$ remains highly elevated compared to isolated disks and during the interval before second passage.

\begin{figure}
\centering
  \includegraphics[width=0.5\textwidth]{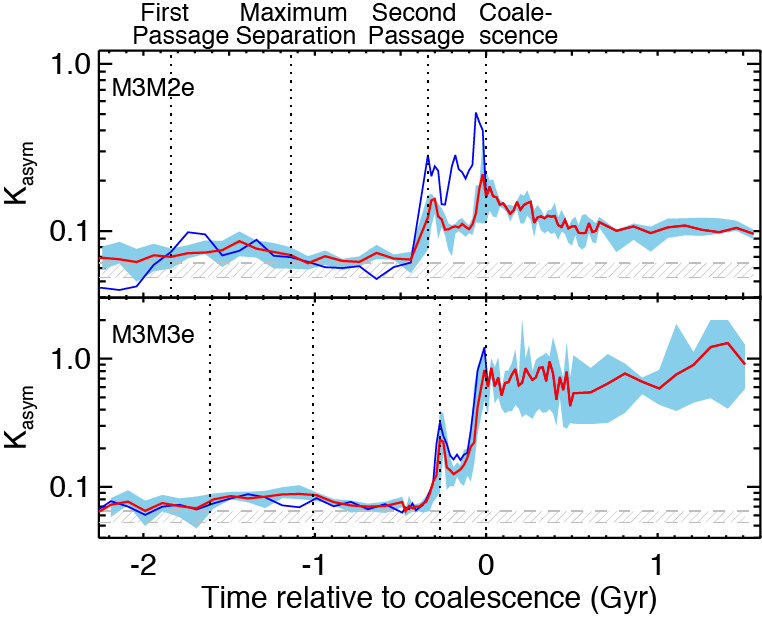} 
\caption{The evolution of $K_{asym}$ traced by stellar populations during the M3M2e and M3M3e simulations.
The dotted lines indicate the stages of interaction: first passage, maximum separation, second passage, and final coalescence.
The red solid lines show the median $K_{asym}$ curve of one galaxy (the major galaxy in the M3M2e simulations) derived from the values of seven viewing angles.
The blue area indicates the distribution of 70\% of the data (values of five viewing angles closet to the median values).
The solid blue line indicates the median $K_{asym}$ curve of the companion galaxy (the minor galaxy in the M3M2e simulations).
The gray shaded area indicate the distribution of 70\% of the values from the isolated M3 galaxy simulations.
} 
\label{fig:kasym_sdb}
\end{figure}

\section{Discussion}
\subsection{Implications to the measurements of galaxy merger rates and merger fractions}
One important application of the large IFS surveys is to constrain the merger abundance of star-forming galaxies using kinematically identified close pairs \citep[e.g.,][]{Lopez-Sanjuan2013} or signatures of complex dynamics \citep[e.g.,][]{Yang2008}.
Our work shows that when defining mergers as galaxies with significantly elevated $K_{asym}$, the MOTs are typically 0.2$-$0.4 Gyr except the equal mass merger with SMG-type progenitors, which has MOT that is approximately twice as long as those of $z\sim0$ mergers due to its more gradual decline in SFR after black hole coalescence.
The MOTs can be shorter if the resolution of kinematic maps is worse than $\sim0.5$ kpc.
Since no noise is added in the kinematic maps, the observable times derived here are likely represent the best case scenario at least with currently achievable resolutions.
Even during the strong interaction phase (i.e. after second passage and before coalescence), only $\sim40-80\%$ of galaxy mergers show significant enhancement in $K_{asym}$ (Figure~\ref{fig:merger_kin}).
The short merger observable times and the incompleteness of merger fractions reinforce the need of careful corrections when deriving galaxy merger rates and merger fractions using kinematic diagnostics.

The merger observable times based on $K_{asym}$ are comparable to the morphologically identified merger observable times using $Gini$ coefficient, $A$, and $M_{20}$ \citep{Lotz2008,Lotz2010}, in which both morphology and kinematics-based identifications are most sensitive to galaxy mergers during the strong interaction phases.
An advantage of kinematic diagnostics is that the complex kinematics are visible for up to several hundred Myr after black hole coalescence (e.g., M3M3e, M3M2h, b6b6e).
Combining morphological and kinematic information can thus provide a more accurate assessment of galaxies' dynamical status.
For instance, when defining galaxies as mergers with either elevated $A$ or $K_{asym}$, the MOTs of M3M2e and M3M3e simulations increase from $0.22$ and $0.36$ Gyr to $0.28$ and $0.38$ Gyr, respectively.

\subsection{Measurements of disk properties}
A key result from recent studies of galaxy kinematics is that the velocity dispersion of disk galaxies are systematically higher at higher $z$ \citep[e.g.,][although local LIRG-type isolated disks typically have higher velocity dispersion as well; \citealp{Bellocchi2013}]{Law2009,Epinat2012,Kassin2012}.
The increased velocity dispersions are often attributed to the enhanced gas fractions in the high$-z$ disk galaxies, which can lead to highly unstable and turbulent dynamics \citep[e.g.,][]{Genzel2011}.
However, given the short merger observable times and the $<100\%$ merger recovery rates (Figure~\ref{fig:merger_kin}) based on the disturbance in kinematics, some of the disk galaxies identified by IFS surveys may be misidentified or a result of mergers.
It is therefore important to quantify the evolution of velocity dispersions during galaxy interactions.

We define a sample of ``disk galaxies'' from the M3M2e simulations (original and doubled gas fractions) as those galaxies having $K_{asym}$ consistent with the $K_{asym}$ of 95\% of the isolated M3 simulations.
The M3M2e simulations are chosen because the disk structure survives after the coalescence.
We measure the intrinsic velocity dispersion\footnote{Here we compare $\sigma_0$ at different interaction stage that are derived in a consistent methodology. Note that these numbers are not necessarily comparable to those in the literature as different groups use varying methods to calculate $\sigma_0$ \citep[e.g., see the discussion in][]{Glazebrook2013,Wisnioski2015}.} ($\sigma_0$) of this disk sample, in which we define $\sigma_0$ as the velocity dispersion at the positions with the largest velocities along the axis of the steepest velocity gradient.
Figure~\ref{fig:disk_s0} shows $\sigma_0$ for the disk sample as a  function of interaction stage.
The star-forming disks surviving after coalescence have a median $\sigma_0$ $\sim4$ times higher than the progenitor disks before the first passage.
Even during the strong interaction phases when the dynamics of star-forming gas is dominated by the bulk motion of two nuclei but not coherent rotation, the measured $\sigma_0$ can be significantly higher than during earlier interaction stages.
This implies that if the disk sample identified by the IFS surveys contains misidentified mergers or merger remnants, this population may also lead to high $\sigma_0$.

\subsection{Limitations of this work}
Unlike optical imaging surveys, kinematic studies based on IFS observations often require pre-selection of the observed samples (e.g., optical and near-infrared colors) and this may introduce biases when converting the observed merger fractions to the overall galaxy merger rates.
To obtain merger recovery rates for arbitrary sample selections, it is important to expand the kinematic analysis conducted in this work to large binary merger simulation library or cosmological simulations that provide a means to test various sample selections that mimic those used in the IFS surveys.
However, the paucity of strong merger-induced starbursts in state-of-the-art large-volume cosmological simulations \citep{Sparre2015} suggests that such simulations may not yet sufficiently resolve the nuclear regions of galaxy mergers. 
High resolution zoom-in cosmological simulations \citep[e.g.,][]{Hopkins2014} can partially overcome this drawback, but they are computationally expensive, making it challenging to assemble a large sample of interacting galaxy simulations with this technique.
Consequently, suites of idealized merger simulations will likely remain the best tool for studies such as the present one for some time.

Although we attempt to address the applicability of our results to $z\gtrsim2$ IFS studies by using the progenitors of gas-rich disks, and the SMG-type progenitors, a possible caveat is that the gas properties assumed in our hydrodynamic simulations may not be comparable to those of high$-z$ star-forming galaxies.
For instance, \citet{Bournaud2011} show that interactions between clumpy disks can lead to more chaotic kinematics compared to the progenitors with stabilized ISM.
However, it is unclear whether the drastic differences shown by the entire gas content \citep[Figure 3 \& 4 in][]{Bournaud2011} are visible with only dense, star-forming gas.
We perform a test run of M3M2e and M3M3e simulations with extreme initial gas fraction (0.8) and soft effective equation of state ($q_{\rm EOS}=0.05$), in which these parameters can lead to highly unstable disk within several hundreds Myr after the start of the simulations and large star-forming clumps similar to some $z\sim1-3$ star-forming galaxies \citep{Springel2005c}. 
Yet without a continuous replenishment of gas in these simulations, the gas fractions decrease to only $0.2-0.3$ during the strong interaction phase and thus galaxy kinematics at this stage is consistent with other simulation runs with lower initial gas fractions.

Finally, we use gas particles with SFR$>0$ (i.e. $n\gtrsim0.1$ cm$^{-3}$) as a proxy of star-forming gas throughout this analysis, yet such simple approximation does not consider possible impacts from dust attenuations or optical depth.
Future implementations of radiative transfer codes such as \sunrise \citep{Jonsson2010} to kinematic analysis will allow us to explore the effects of dust extinctions.
The mock IFS datacubes will also allow us to intuitively include observational effects such as skylines in the near-infrared observations.

\begin{figure}
\centering
  \includegraphics[width=0.5\textwidth]{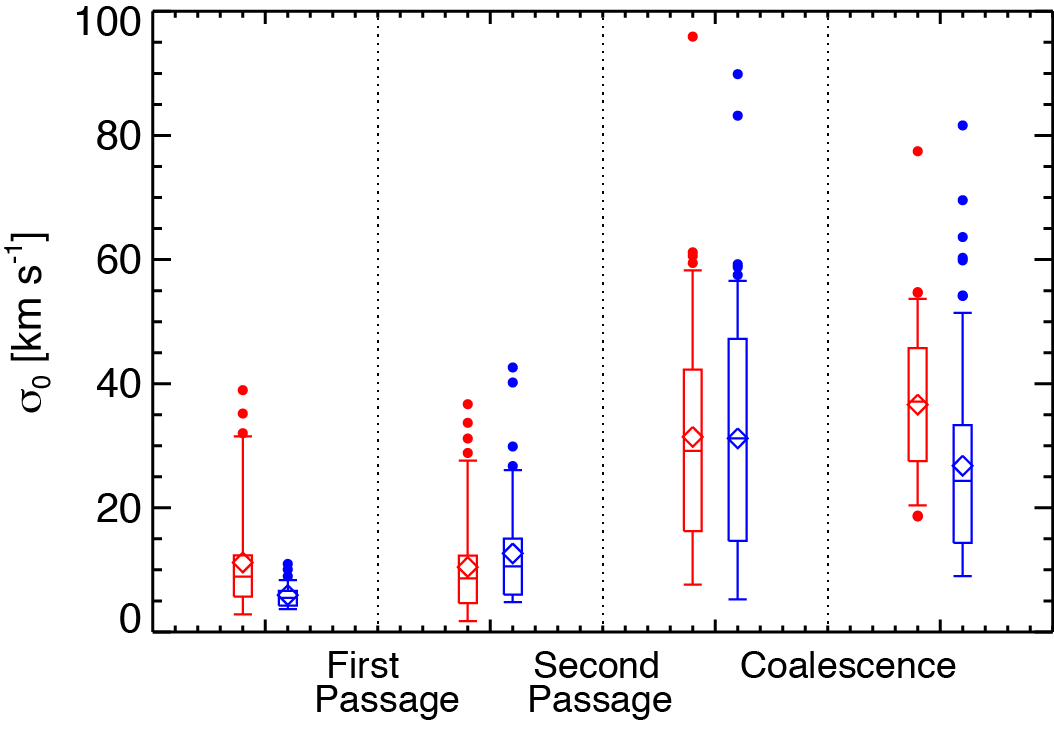} 
\caption{Intrinsic velocity dispersion ($\sigma_0$) of ``disk galaxies,'' i.e., interacting systems with $K_{asym}$ consistent with isolated disks (the M3 simulations). 
From left to right, the Figure shows the intervals from: the start of the interactions to first passage, from first passage to second passage, from second passage to coalescence, and after coalescence.  
Results for the standard M3M2e simulations are shown in red, and those for the gas-rich M3M2e simulations are shown in blue. 
The boxes indicate the locus of 70\% of the measurements, and the vertical bars encompass 95\% of the data. 
The filled dots are outliers that fall outside the central 95\% of the points. 
The horizontal bars in the boxes indicate the median values and the diamonds indicate the means. 
} 
\label{fig:disk_s0}
\end{figure}


\section{Conclusions}

We study the dynamics of star-forming gas in interacting galaxies using a set of hydrodynamic simulations with stellar mass ratios of 1:1 and 1:4.
Using the SPH gas particles with SFR$>0$ as a proxy for star-forming gas, we construct two-dimensional velocity and velocity dispersion maps throughout the interaction sequence.
We quantify the disturbance in the kinematic maps based on the measurements of kinematic asymmetries ($K_{asym}$), and we define galaxies as observable mergers when their $K_{asym}$ is significantly elevated above the values of isolated disk galaxies.
Our conclusions are summarized as follows:

\begin{enumerate}

\item The evolution of $K_{asym}$ mirrors that of the morphological asymmetries ($A$) in both equal and unequal mass galaxy mergers (our M3M3e and M3M2e simulations), in which they most significantly deviate from the isolated disk simulations during the strong interaction stage.

\item When defining mergers as snapshots having $K_{asym}$ higher than 95\% of the isolated disk simulations, the merger observable time (i.e. the time duration that merger signatures are detectable) are 0.22$\pm$0.04 for the M3M2e simulations and 0.36$\pm$0.06 for the M3M3e simulations. 
These observable times are typically $0.2-0.4$ Gyr based on simulations with various orbital parameters, initial disk orientations, and gas fractions.

\item The 1:1 and 1:4 galaxy mergers with SMG-type progenitors (our b6b6e and b6b5e simulations) show a similar evolution in $K_{asym}$ as the $z\sim0$ mergers, in which $K_{asym}$ only begins to elevate significantly after the second passage.
However, the merger observable time of b6b6e is approximately twice longer than M3M3e because the SFR of b6b6e declines more gradually than M3M3e after black hole coalescence.

\item The merger observable time are sensitive to the spatial resolution used to construct the kinematic maps. 
In our test with the M3M2e simulations, the observable time decrease from 0.22 Gyr to 0.14 Gyr when using 1 kpc $\times$ 1 kpc instead of 0.5 kpc $\times$ 0.5 kpc grids.

\item We find that the merger observable probability shows a strong trend with the interaction stage. 
The measured merger recovery rates are typically below 20\% before second passage.
The recovery rates increase to $40-80\%$ during the strong interaction stages, and the scatter is even larger after the black hole coalescence depending on whether the disk structures survive during interactions.

\item We derive the intrinsic velocity dispersion ($\sigma_0$) of galaxies consistent with isolated disks (in $K_{asym}$) for the M3M2e simulations.
We find that the surviving disks after coalescence have a median $\sigma_0$ $\sim4$ times higher than the progenitor disks.
The enhanced $\sigma_0$ is also measured during the strong interaction phases even when the systems are not indeed rotating disks.

\end{enumerate}

\acknowledgments
C-LH, HAS, MLNA, and JRM-G wish to acknowledge partial funding support from NASA grants NNX14AJ61G and NNX15AE56G.
CCH is grateful to the Gordon and Betty Moore Foundation for financial support.
The computations in this paper were run on the Odyssey cluster supported by the FAS Division of Science, Research Computing Group at Harvard University.


\bibliographystyle{../Reference_Bib/apj}
\bibliography{../Reference_Bib/Cosmos}

\end{document}
